\documentclass[twocolumn,a4paper,superscriptaddress,floatfix,showpacs]{revtex4}
\usepackage{amsmath,amssymb,epsfig,color,textcase}

\begin{document}

\title{Interplay between lattice, orbital, and magnetic degrees of freedom in
the chain-polymer Cu(II) breathing crystals.}
\author{S.V.~Streltsov}
\affiliation{Institute of Metal Physics, S.Kovalevskoy St. 18, 620990 Ekaterinburg, Russia}
\affiliation{Ural Federal University, Mira St. 19, 620002 Ekaterinburg, Russia}
\email{streltsov@imp.uran.ru}

\author{M.V. Petrova}
\affiliation{International Tomography Center SB RAS, Institutskaya str. 3a, Novosibirsk, Russia}

\author{V.A. Morozov}
\affiliation{International Tomography Center SB RAS, Institutskaya str. 3a, Novosibirsk, Russia}

\author{G.V. Romanenko}
\affiliation{International Tomography Center SB RAS, Institutskaya str. 3a, Novosibirsk, Russia}

\author{V.I. Anisimov}
\affiliation{Institute of Metal Physics, S.Kovalevskoy St. 18, 620990 Ekaterinburg, Russia}
\affiliation{Ural Federal University, Mira St. 19, 620002 Ekaterinburg, Russia}

\author{N.N. Lukzen}
\affiliation{International Tomography Center SB RAS, Institutskaya str. 3a, Novosibirsk, Russia}


\date{\today}

\begin{abstract}
The chain-polymer Cu(II) ``breathing crystals'' C$_{21}$H$_{19}$CuF$_{12}$N$_4$O$_6$
were studied using the X-ray diffraction and {\it ab initio} band structure
calculations. We show that the crystal structure modification at T=146 K, associated
with the spin crossover transition, induces the changes of the
orbital order in half of the Cu sites. This
in turn results in the switch of the magnetic interaction sign in
accordance with the Goodenough-Kanamori-Andersen theory of the
coupling between the orbital and spin degrees of freedom.
\end{abstract}

\maketitle

\section{Introduction \label{intro}}
The conventional phenomenon of spin crossover (SCO) is a well-known change of
the spin state observed in some octahedral coordinated transition metal
complexes.~\cite{Gutlich-2004}
There are exist thermally, doping, photo- and pressure induced
SCO transitions.~\cite{Gutlich-2004,Kugel-12,Ogawa-00,Shepherd-11}.
In the classical case of Fe(II) complexes with $d^6$ electron configuration the thermal
SCO involves a transition from the low spin state (S=0, $t_{2g}^6$) to the
high spin state (S=2, $t_{2g}^4 e_g^2$) \cite{Hauser-2004,Streltsov2011} at elevating temperature.
The SCO compounds represent bright examples of a bistability in the molecular
crystals\cite{Cobo-2008} and are promising candidates for multifunctional
materials\cite{Gutlich-1994} with potential applications to the memory devices,
the optical, temperature and pressure sensors etc.~\cite{Letard-2004}

Recently rather different SCO transitions were found in the
chain-polymer compounds Cu(II) with nitroxides.~\cite{Ovcharenko-2002}
These compounds contain chains of the exchange clusters  with two or three
spins. The exchange clusters contain Cu$^{2+}$, ligands and some
organic radicals. For the essential structural changes
in the polyhedral surrounding Cu ions under the SCO transitions
these compounds were called ``breathing crystals''.~\cite{Ovcharenko-2004}
The main feature of the ``breathing crystals'' compounds is their ability to pass through
reversible thermal induced structural transformations (often similar to phase transitions)
accompanied by changes of magnetic susceptibility and optical
properties.~\cite{Tretyakov-2012}.
Note that one should not confuse the ``breathing crystal'' with the well known dynamical
``breathing mode'' -- \, a specific collective excitation of confined systems of quantum \cite{Bonitz-2012}
or classical \cite{Olivetti-2009} particles.

The classical SCO transition, associated with the change of spin state of single ion, is
impossible for isolated $d^9$ centers of Cu(II) (S=1/2, $t_{2g}^6 e_g^3$).
Thus, the reason of unusual spin transitions inherent in the
Cu(II) complexes with nitroxides possibly arise from the change of
the total electron spin of a whole exchange cluster.

Magnetic measurements show that the temperature induced SCO transition
in one of the Cu(II) breathing crystals, characterized by the chemical formula
C$_{21}$H$_{19}$CuF$_{12}$N$_4$O$_6$, is accompanied by the lost of the half
of local spins.~\cite{Ovcharenko-2004} This fact can be explained by the
formation of spin singlets (S=0) in the half of the exchange clusters,
but the reason for this is unknown.

The aim of our paper is to provide a microscopic description
of the changes in the magnetic properties of C$_{21}$H$_{19}$CuF$_{12}$N$_4$O$_6$,
often abbreviated as Cu(hfac)$_2$ L$^{Me}$ in the chemical literature.
With the use of the density functional
theory (DFT) we found that there is an interplay between the magnetic
properties, orbital structure and lattice distortions in the
 ``breathing crystals''. These correlations between different degrees of freedom
results in the SCO transition at 146~K in the compound under consideration.

\section{Crystal structure \label{SEC:crysstr}}
The crystal structure of C$_{21}$H$_{19}$CuF$_{12}$N$_4$O$_6$ was solved
from the X-ray single crystals diffraction data. The data were collected
using a SMARTAPEXCCD (Bruker AXS) automated diffractometer with aHelix
(OxfordCryosystems) open-flow helium cooler using the standard procedure
(Mo K $\alpha$ radiation).
The structures were solved by direct methods and refined by the full-matrix
least-squares procedure anisotropically for non-hydrogen atoms. The H atoms were
partially located in difference electron density syntheses, and the others were
calculated geometrically and included in the refinement as riding groups.
All calculations were fulfilled with the SHELXTL 6.14 program package.

\begin{figure}[b!]
 \centering
 \includegraphics[clip=false,width=0.5\textwidth]{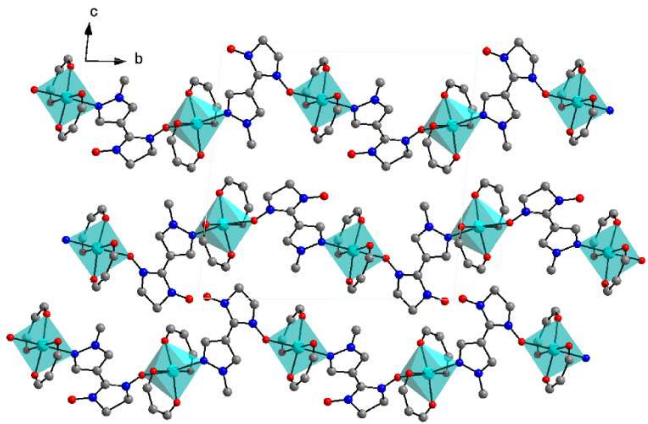}
\caption{\label{Fig1_Moroz}. (Color online) The crystal structure of high temperature
triclinic phase of the ``breathing crystal'' compound C$_{21}$H$_{19}$CuF$_{12}$N$_4$O$_6$.
Turquoise balls are the Cu ions, red and blue balls are the O and N atoms.
Grey balls are the C atoms. The fluorine atoms and the hydrogen atoms are omitted for
clarity of the figure. Coordination units CuO$_5$N are marked with turquoise octahedra.}
\end{figure}

Crystal data for the compound  Cu(hfac)$_2$L$^{Me}$ are the following:
C$_{21}$H$_{19}$CuF$_{12}$N$_4$O$_6$, FW = 714.94, T = 240/110 K, triclinic crystals,
space group P-1,
a=12.1987(9)/ 11.9560(9)$\AA$,
b=15.5950(11)/ 15.0506(12)$\AA$,
c=15.8716(11)/ 15.8657(12)$\AA$,
$\alpha$ = 84.459(2)/81.306(1)$^\circ$,
$\beta$ = 74.132(2)/76.943(2)$^\circ$,
$\gamma$ = 87.315(1)/85.502(1)$^\circ$,
V = 2890.1(4)/2746.3(4)$\AA^3$ , Z=4,
D$_{calc}$ = 1.643 /1.729 g/cm$^3$ ,
$\mu$ = 0.875 /0.921 mm$^{–1}$,
12570/11868 measured reflections
($\theta_{max}$ = 23.35/23.30$^\circ$),
8291/7907 unique reflections
(R$_{int}$ = 0.0293/0.0230), 6043/6521 reflections with
$I > 2\sigma_I$, 793/794  refined parameters;
GOOF = 1.061/1.103, $R1$ = 0.0598/0.0481, $wR2$ = 0.1500/0.1307  ($I > 2\sigma_I$).
The list of the atomic positions for T=240~K  and T=110~K
is given in the Supplemental Materials.~\cite{Streltsov12sup}

\begin{figure}[t!]
 \centering
 \includegraphics[clip=false,width=0.5\textwidth]{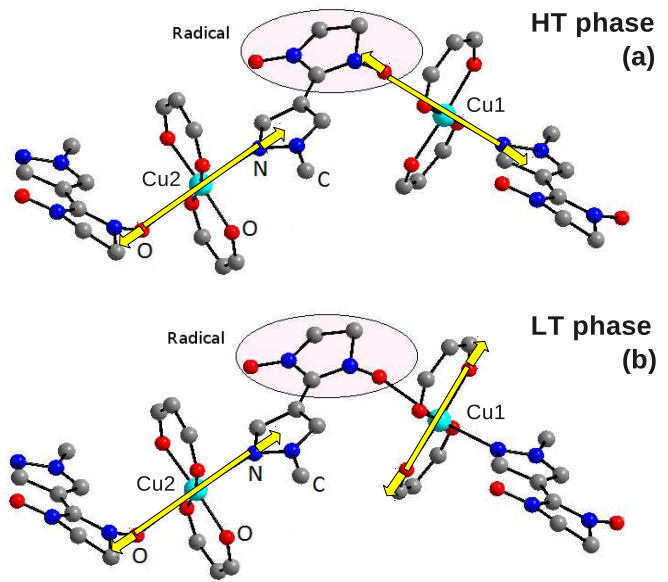}
\caption{\label{Fig2_Moroz}. (Color online) The structure of isolated polymer chain of
the ``breathing crystal'' compound C$_{21}$H$_{19}$CuF$_{12}$N$_4$O$_6$.
The color coding is the same as in Fig.~\ref{Fig1_Moroz}
The fluorine atoms and the hydrogen atoms are omitted for clarity of the figure.
Methyl ligand characterizing this compound is marked by C. In other compounds of the
``breathing crystals'' family the ligand may be propyl or butyl type.
A fragment of nitroxide radical with nonzero spin densities on atoms is marked
with the oval. The direction of the CuO$_5$N octahedra elongation are shown by
the yellow bars and arrows.}
\end{figure}

The crystal structure of C$_{21}$H$_{19}$CuF$_{12}$N$_4$O$_6$ consists of the polymer
chains  running along $b$ direction (Fig.~\ref{Fig1_Moroz}) with a ``head-to-tail''
motif containing
the exchange clusters of type N-$\dot{\texttt{O}}$-Cu$^{2+}$. An isolated chain of the
exchange clusters is presented in Fig.~\ref{Fig2_Moroz} with the F and H
atoms being omitted for simplicity.

At high temperature (HT) phase the CuO$_5$N units are the elongated octahedra with Cu-O axial
distance of $\sim$2.5 \AA~ and Cu-N distance of $\sim$2.3 \AA. Equatorial
distances Cu-O are about 1.95 \AA. With decrease of the temperature the compound undergoes
the SCO transition in the vicinity of 150 K, which is accompanied by the
substantial structural changes. As a result of the transition within a half of
the CuO$_5$N octahedra the elongation direction changes. In the octahedra
surrounding Cu1 ion two equatorial bonds with oxygens turn out to be the longest, while
in the Cu2O$_5$N octahedra the elongation direction still coincides
with the O-Cu-N bond, like in the HT phase. Corresponding bond lengths
can be found in Tab.~\ref{tab:1} for the HT (240 K) and
LT (110 K) phases.

\section{Previous calculations of the ``breathing crystals''}
The first attempt to understand a nature of the spin exchange in the clusters containing
the Cu atoms and a stable nitroxyl radical was performed rather far ago by Musin
et al.~\cite{Musin-1992} The authors provided a detail quantum-chemical analysis of
the possible mechanisms of the exchange interaction in the magnetic fragments
$\dot{\texttt{Cu}}$(II)$\cdots$O-$\dot{\texttt{N}}<$
(or $>\dot{\texttt{N}}$-O$\cdots$ $\dot{\texttt{Cu}}$(II) $\cdots$O-$\dot{\texttt{N}}<$) of bischelating
complexes of Cu(II) with nitroxyl radicals. The drawback of this and some
other \cite{Gorelik-2008, Zueva-2009, Neese-2009, Gritsan-2010} treatments was in the consideration of an
isolated fragment rather than the crystal as a whole.

Till now the only consistent calculation of the electronic and magnetic properties of
the ``breathing crystals'' was performed in Ref.~\onlinecite{Postnikov-2005},  where the spin densities and
the magnetic moments of a heterospin compound based Cu(II) hexafluoroacetylacetonate
 (Cu(hfac)$_2$; hfac=CF$_3$-C(O)-CH-C(O)-CF$_3$)  in combination with a stable
nitronyl nitroxide radical were calculated.
This system is similar to the one of interest in the present paper, the difference is in another
substitute  in the position 1 of the pyrazol ring -  the ethyl radical instead of the methyl
one in our system. This difference results in another organization of the polymer chain: in
the case considered in Ref.~\onlinecite{onlinePostnikov-2005} the chain of the exchange clusters has
``head-to-head'' coupling of the ligands to the magnetic Cu atoms embedded in
(hfac)$_2$ blocks, while in the present case the chain motif is ``head-to-tail''.

Thus, in Ref.~\onlinecite{Postnikov-2005} the corresponding chain contains ``three spin - isolated spin''
structure, while in the case considered in the present paper the chain is composed of
two spin clusters. This results in a different magnetic properties.
 It should also be mentioned that in Ref.~\onlinecite{Postnikov-2005} only a high temperature phase of the crystal
was calculated. At the same time both high-temperature and especially low temperature phase
represent a great interest for study of phase transitions.

\section{Calculation details}
The pseudo-potential PW-SCF code was used for the band structure
calculations.~\cite{Giannozzi2009}  We utilized ultrasoft pseudo-potentials
with nonlinear core correction (for better description
of the magnetic interactions) with Perdew-Burke-Ernzerhof
(PBE) version of the exchange-correlation potential.~\cite{Perdew1996}
The charge density and kinetic energy cut-offs equal 35 Ry and
180 Ry, respectively. The integration in the k-space in the course of the
self-consistency was performed over the mesh of 4 k-points in the
Brillouin zone using gaussian smearing of 13.6 meV. The density of states
was calculated with the smearing of 40.8 meV.

The correlations on Cu sites were taken into account within the frameworks
of the GGA+U approximation (generalized gradient approximation with account
of on-site Coulomb repulsion).~\cite{Anisimov1997} The intra-atomic exchange interaction
J$_H$ and on-site Coulomb repulsion parameters for Cu$^{2+}$ ions
were chosen to be 0.9 and 7.0 eV respectively~\cite{Streltsov12Cu,Liechtenstein1995}.

\begin{table}
  \caption{Bond distances of Cu2O$_5$N octahedra in $\AA$. The first four
rows correspond to distances in the equatorial plane,
i.e. perpendicular to the chain direction, while the last two ones
to the axial direction (along the chain). The
space group is the same P-1 for both phases.}
  \label{tab:1}
  \begin{tabular}{c|cc|c|cc}
    \hline
    \hline
Bond & HT phase & LT phase & Bond & HT phase & LT phase\\
     & T=240 K  & T=110 K  &      & T=240 K  & T=110 K \\
    \hline
Cu1-O  & 1.95 & 1.97  & Cu2-O  & 1.93 & 1.85\\
Cu1-O  & 1.95 & 2.16  & Cu2-O  & 1.95 & 1.85\\
Cu1-O  & 1.96 & 1.98  & Cu2-O  & 1.96 & 1.97\\
Cu1-O  & 1.97 & 2.23  & Cu2-O  & 1.97 & 1.97\\
    \hline
Cu1-O  & 2.49 & 1.99  & Cu2-O  & 2.49 & 2.42\\
Cu1-N   & 2.30 & 2.02 & Cu2-N  & 2.32 & 2.32\\
    \hline
    \hline
  \end{tabular}
\end{table}

The calculations were performed for the experimentally measured crystal structure,
presented in Ref.~\onlinecite{Streltsov12sup}
However, in order to decrease the number of atoms in the
unit cell (from 252 to 126) we used P1 instead of the P-1 space group.
The inversion center in the P-1 space group produces additional
chains, so that reducing the crystal symmetry we neglect the interchain
interactions. This seems to be a reasonable approximation, since
the main changes in the crystal structures at the SCO transition
occur within a chain.

\section{Calculation results}
\begin{table}
  \caption{The results of the GGA+U calculation for two different
crystal structures corresponding to the temperatures T=110 K (LT)
and T=240 K (HT) and to the FM and AFM types of magnetic order of the Cu spins. All the values are in
$\mu_B$ units. The total and absolute (abs) magnetization are per unit cell.}
  \label{tab:2}
  \begin{tabular}{ccccc}
    \hline
    \hline
            & HT phase & HT phase & LT phase & LT phase\\
            & FM       & AFM      &     FM   &  AFM \\
    \hline
total magn. & 3.92 & 0.06 & 2.00 & 2.00 \\
abs   magn. & 4.40 & 4.44 & 4.07 & 4.07 \\
moment Cu1  & 0.52 & -0.52 & 0.45 & -0.45 \\
moment Cu2  & 0.55 & 0.55 & 0.55 & 0.55\\
    \hline
    \hline
  \end{tabular}
\end{table}

\subsection{High-temperature phase}
We start with the calculations of the HT phase. The compound under
consideration experimentally is known to be paramagnetic.~\cite{Ovcharenko-2004}
However, in any band structure calculation the translational symmetry is assumed, so
that one may calculate the ferromagnetic (FM), when both Cu ions in the
unit cell (u.c.) have the same spin direction, antiferromagnetic,
with the opposite spins on the neighboring Cu atoms, or non-magnetic,
when up and down spins on all ions are equally populated. It's clear
that the non-magnetic configuration is the worst approximation, since
there must be local magnetic moments in the paramagnetic insulator.
In the following we will investigate the FM and AFM solutions,
which actually provide very similar results since magnetic ions (Cu)
are quite far a way from each other.

\begin{figure}[t!]
 \centering
 \includegraphics[clip=false,angle=270,width=0.45\textwidth]{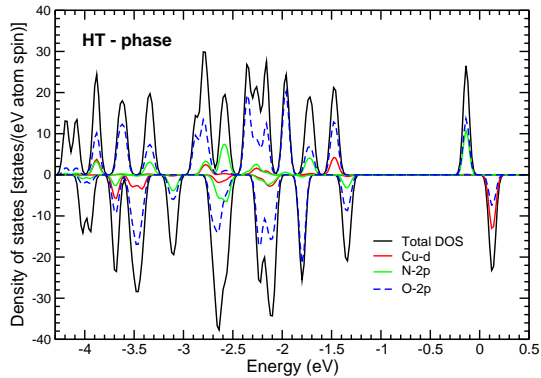}
\caption{\label{TDOS-HS}. Results of the GGA+U calculation for the high temperature phase,
where the spins on different Cu ions are ferromagnetically ordered.
Positive (negative) values correspond to the
spin up (down). The gaussian smearing 0.04 eV was used.
The Fermi energy is set to zero.}
\end{figure}

Indeed, the absolute values of the magnetic moments on two Cu ions
were found to be 0.52 $\mu_B$ and 0.55 $\mu_B$, the same for the FM and
AFM solutions (see Tab.~\ref{tab:2}). They differ from 1 $\mu_B$
expected for the isolated Cu$^{2+}$ ions due to a strong hybridization
with ligands and formation of a molecular orbital
on which a single hole in the $3d$ shell of the Cu$^{2+}$ ion is localized.
This molecular orbital (or the Wannier orbital) has the contributions
of the Cu $d$ and ligand $p$ orbitals, so that the spin moment on the
whole molecular orbital should be 1 $\mu_B$, but the part of the spin
density projected on the Cu $3d$ states provides only a part of it.
This is clearly seen from the density of states plot presented in
Fig.~\ref{TDOS-HS}. The peak at $\sim 0.1$ eV corresponding
to a single hole in the Cu $3d$ shell shows significant contribution
coming from the O $2p$ states. Another illustration of the considerable
mixing between Cu $3d$ and O $2p$ states can be found in
Figs.~\ref{spin-density-Cu1} and \ref{spin-density-Cu2}, where the spin density
plot for the LT phase is presented.

It is useful to proceed with the analysis of the total magnetization
per unit cell defined as $m_{tot} = \sum_i  m_i$, where
$m_i=g \mu_B s_i$ is the
magnetization on the $i$th atom, and $s_i$ -- its spin moment.
The total magnetization equals 3.92$\mu_B$ in the FM and
0.06$\mu_B$ in the AFM configuration. Since there are two pairs
radical-Cu in the unit cell, it means that at least in the FM
configuration the spins on the radicals are {\it parallel} to the spin
moment of the molecular orbital on the neighboring (to this radical)
Cu ion. The spin density projection shows that the largest
moment on $p-$elements (O, C, F, and N) are indeed
parallel to the moments of Cu.  The values of the
largest moments ($>$0.1$\mu_B$) in the case of the FM order
are the following: O10 (0.24 $\mu_B$); O11 (0.30 $\mu_B$);
O12 (0.31 $\mu_B$); N1,N2,N5 (0.22 $\mu_B$);
N6 (0.22 $\mu_B$).

In other words the pair radical-Cu is in the
triplet ground state (S=1, where S is the spin of the pair).
The deviation from the 4~$\mu_B$ for the FM and from 0~$\mu_B$ for
the AFM solutions is attributed to the sparse mesh in the
$k-$space, used to integrate energy bands in such a large unit cell,
consisting of 126 atoms.

In principle one may explain the ferromagnetic coupling (parallel
spin arrangement) of the spins on the Cu ions and radicals in two ways.
First of all one may argue that the local spin on Cu simply magnetizes all the
surrounding ions. Microscopically this means that due to a strong
hybridization the Zeeman (spin) splitting in the Cu $3d$ shell
spreads out on the $s$ and $p$ shells of the neighboring ligands.
However, this picture is too simplified and doesn't take
into account the details of the electronic structure of the
compound under consideration and is unable to explain the
antiferromagnetic coupling between Cu and the radical which is
observed at low temperatures and will be discussed latter.
Therefore bellow we present the model, which explains
in details how the magnetic coupling with the radicals
is related to the local lattice distortions and the orbital
structure of the Cu$-3d$ shell.
\begin{figure}[t!]
 \centering
 \includegraphics[clip=false,width=0.45\textwidth]{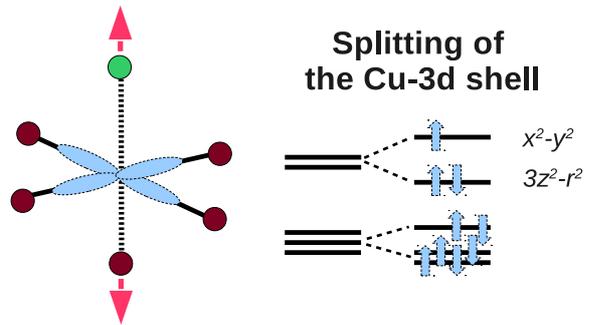}
\caption{\label{Octa-HT}. (Color online) The sketch of the local
distortion of CuO$_5$N octahedra (oxygen and nitrogen atoms are
brown and green balls, respectively) in the high temperature phase.
The direction of the distortion is shown by the red arrows.
The elongation of the octahedra leads to the
stabilization of a hole on the $x^2-y^2$ orbital.}
\end{figure}

According to the crystal structure analysis
presented above in Sec.~\ref{SEC:crysstr} the distortions of both CuO$_5$N octahedra (which
belong to the same unit cell) are quite similar. Since the Cu$^{2+}$ ion is Jahn-Teller
active, both octahedra are strongly distorted. They are elongated in
the direction of the O-Cu-N bond. The average Cu-O distance in the equatorial
plane is $\sim$1.95\AA, while the bond lengths
with the apical ligands are $\sim$2.30 and 2.49 \AA, for the
Cu-O and Cu-N bonds respectively.

Such a distortion of the local surrounding of the Cu$^{2+}$
leads to a certain splitting in the $e_g$ shell of these ions:
the orbital of the $x^2-y^2$ symmetry turns out to be higher in energy
than $3z^2-r^2$, as it is shown in Fig.~\ref{Octa-HT}.
As a result the hole localizes on this $x^2-y^2$ molecular orbital, which lies
in the plane orthogonal to the bond with the radical. By symmetry
this orbital may hybridize only with the O $2p$ states, not with the
N $2p$ orbitals. This is clearly seen in Fig.~\ref{TDOS-HS}, where the peak
corresponding to the hole states at $\sim$0.1 eV, does not have the contribution
coming from the N $2$p states.

Thus the overlap between magneto-active orbital centered on the Cu$^{2+}$
ion and the molecular orbital bearing the local spin on
the radical is negligible. The only possible magnetic coupling
between Cu and the radical is via orbital of the $3z^2-r^2$ symmetry.
But this interaction between the completely filled $3z^2-r^2$ orbital
and the partially filled radical molecular orbital must be
ferromagnetic according to famous Goodenough-Kanamori-Andersen (GKA)
rules.~\cite{Goodenough1963} This is
exactly what we observe in the calculation.

\subsection{Low-temperature phase}
The situation in the low temperature (LT) phase is more complicated, mainly
due to the change in the direction of the elongation in a half of the
CuO$_5$N octahedra. In the LT phase the octahedra surrounding the
Cu1 ions turn out to be elongated not in the direction
of the chain, but perpendicular to it. This is schematically shown in
the Fig.~\ref{Fig2_Moroz}(b). The change of the elongation direction
results in the rotation of the single magneto-active orbital
on the Cu1 ion. The symmetry of this orbital must be the same, $x^2-y^2$,
(here we used the notations of the local coordinate system, where $z$-axis
corresponds to elongation direction), but two lobes of the orbital must
point to the radical.
\begin{figure}[t!]
 \centering
 \includegraphics[clip=false,width=0.5\textwidth]{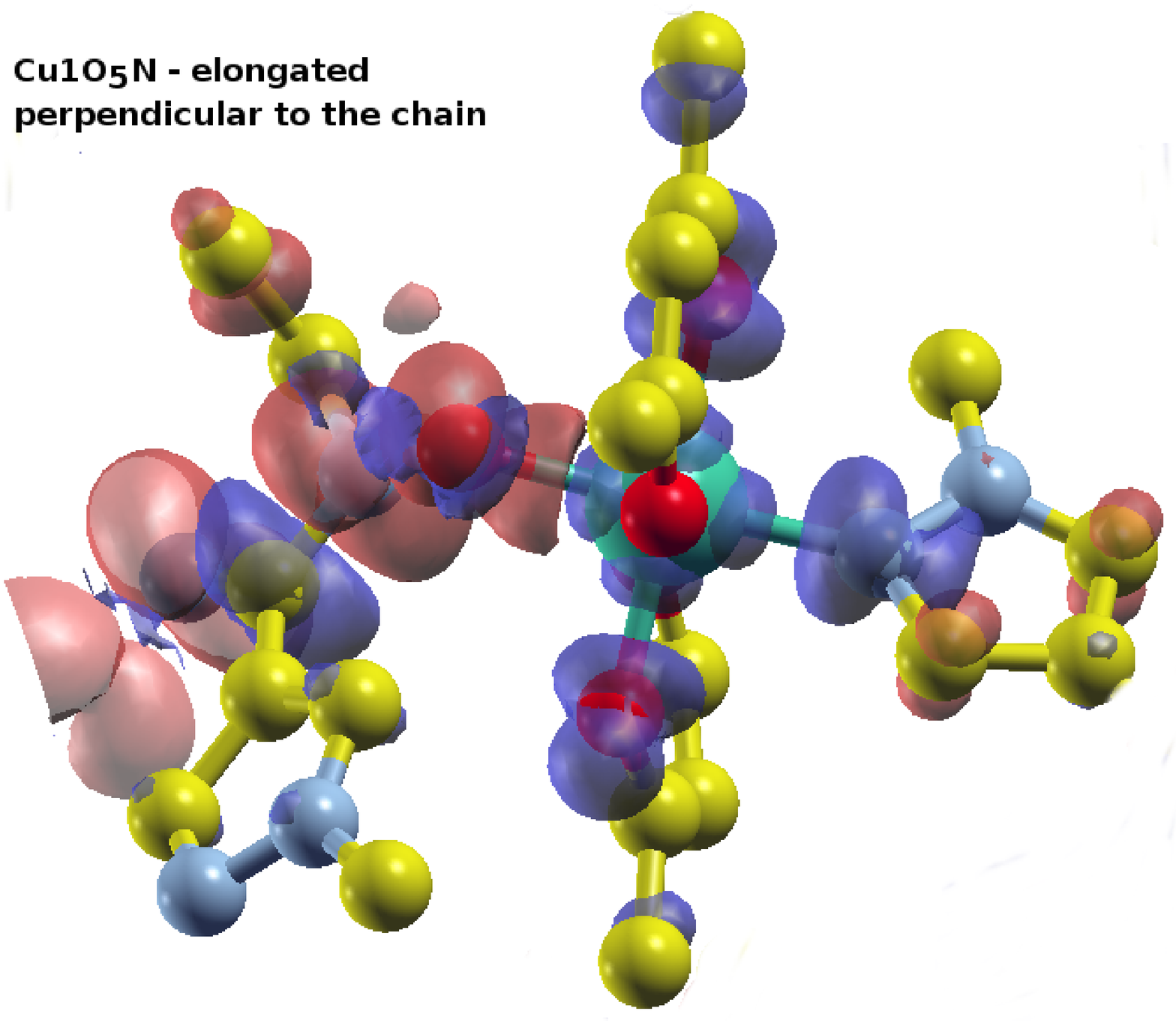}
\caption{\label{spin-density-Cu1}. (Color online) The spin density ($\rho^{\uparrow}(\vec r) -
\rho^{\downarrow}(\vec r)$) in the vicinity of the Cu1 ion
obtained in the GGA+U calculation
for LT phase. The Cu, N, O and C ions are shown as light green, light blue,
red and yellow balls correspondingly. 
The parts of the spin density, which have different signs
are painted by different colors: brown and violet.
Thus, one may easily seen that the spins on the radical and on the Cu1
are antiparallel. The plot is for the FM order of Cu spins,
but the AFM order gives qualitatively the same. }
\end{figure}

In the Fig.~\ref{spin-density-Cu1} the spin density
(difference between the charge densities for two spin projections) around
the Cu1 ion obtained in the GGA+U calculation is shown. In
the case of the Cu$^{2+}$ ($d^9$) this corresponds to the spatial
distribution of the single unoccupied orbital. One may see that as it
was described in details in the previous section the single hole is
actually stabilized not on the atomic, but on the molecular
(Wannier) orbital, which has significant contribution on the neighboring
ligands. The symmetry of the orbital is $x^2-y^2$, but
it is pointed exactly at the spin density centered on the
radical (left part of the Fig.~\ref{spin-density-Cu1}). The strong
overlap between the half-filled orbitals centered on the Cu1 ion
and on the radical results in the strong antiferromagnetic coupling $J \sim 2t^2/U$
according to the GKA rules.~\cite{Goodenough1963} Here $t$ is the
hopping integral and $U$ - the on-site Coulomb repulsion parameter.
The fact that this super-exchange interaction in the real calculation does
lead to the antiferromagetic order of the spins on the radical
and Cu1 is clearly seen from the Fig.~\ref{spin-density-Cu1}.
The signs of the spin density on these two objects are indeed
different.

The presence of the molecular orbital centered on the radical
with the spin antiparallel to the spin on one of the Cu ions
is also seen in Fig.~\ref{TDOS-LT}. In contrast to the HT
phase there are two peaks above the Fermi level. One of them
has contributions coming from the Cu $3d$, N $2p$ and O $2p$ states,
while another only from the N $2p$ and O $2p$ states. The last
one has the spin projection opposite to the spins on the Cu
ions.

\begin{figure}[t!]
 \centering
 \includegraphics[clip=false,width=0.5\textwidth]{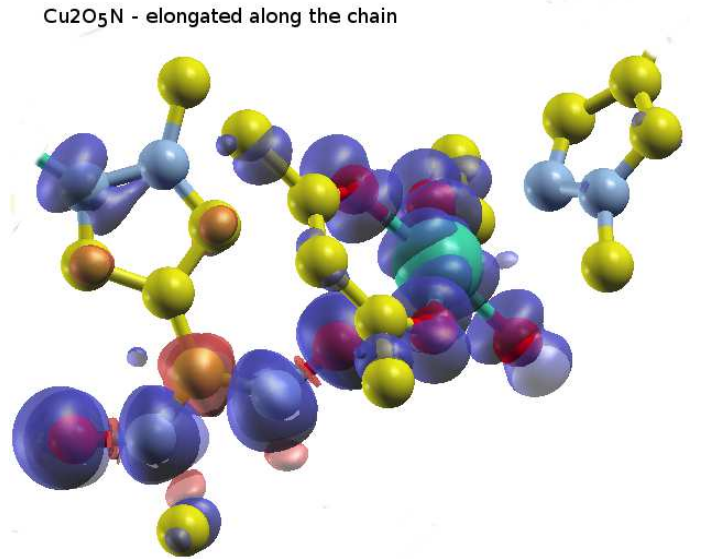}
\caption{\label{spin-density-Cu2}. (Color online) The spin density ($\rho^{\uparrow}
(\vec r) - \rho^{\downarrow} (\vec r)$) in the vicinity of the Cu2 ion
 obtained in the GGA+U calculation for LT phase. 
The Cu, N, O and C ions are shown as light green, light blue,
red and yellow balls correspondingly.
The parts of the spin density, which have different signs
are painted by different colors: brown and violet. Thus, the spins
on the radical and on the Cu2 are parallel.
The plot is for the FM order of Cu spins,
but the AFM order gives qualitatively the same. }
\end{figure}

\begin{figure}[t!]
 \centering
 \includegraphics[clip=false,angle=270,width=0.45\textwidth]{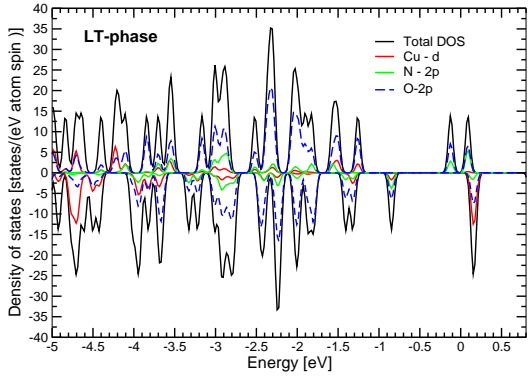}
\caption{\label{TDOS-LT}. Results of the GGA+U calculation for the low temperature phase,
where the spins on different Cu ions are ferromagnetically ordered.
Positive (negative) values correspond to the
spin up (down). The gaussian smearing 0.04 eV was used.
The Fermi energy is set to zero.}
\end{figure}

The spin density centered on the Cu2 ion is presented in Fig.~\ref{spin-density-Cu2}.
One may see that it is concentrated in the plane perpendicular to the
charge density of the radical (so called antiferro-orbital ordering).
This leads to the FM coupling between Cu2 and the neighboring radical,
as in the HT phase according to the GKA rules. Indeed the sign
of the spin density is the same on the radical an on the Cu2 ion.

The fact that there must be a different magnetic coupling in two
pairs Cu-radical in the LT phase according to the
charge density analysis is also seen from the values
of the total and absolute magnetic moments per unit cell.
The total magnetic moment per unit cell is the same, 2.0 $\mu_B$ in both
FM and AFM configurations, which shows that there is one pair of electrons on
the Cu$^{2+}$ ion and the radical with the parallel spin direction
and another one with the antiparallel spins. The fact that
the absolute magnetization defined as $m_{abs} = \sum_i  |m_i|$
is equal to 4.07 $\mu_B$ additionally supports this interpretation.

The absolute values of magnetic moments were found to be the same
in the FM and AFM solutions and equal to 0.45 $\mu_B$ and 0.55 $\mu_B$
for the Cu1 and Cu2 ions respectively (a difference in the values of
magnetic moments as compare to the high temperature phase can be
connected with a slightly different hybridization with ligands,
due to different Cu-O(N) distances).

The strong antiferromagnetic coupling results in the spin singlet
state (S=0) formation in the Cu1 exchange clusters for which the CuO$_5$N octahedra
 are elongated perpendicular to chain direction. This is
exactly what is observed experimentally - the lost of the half of localized
spins in the low temperature phase.~\cite{Ovcharenko-2004}

\section{Conclusions}

We present a  microscopic theory, which describes anomalous
changes of the magnetic properties of
C$_{21}$H$_{19}$CuF$_{12}$N$_4$O$_6$ ``breathing crystal''
through the coupling between the spins and another degrees of freedom
such as orbital and lattice.

With the use of the {\it ab initio} band structure calculations
we show that the change of the crystal structure
of C$_{21}$H$_{19}$CuF$_{12}$N$_4$O$_6$ with decrease of the temperature
results in the rotation of the half-filled orbital of one of the Cu$^{2+}$ ions in the unit cell.
The $x^2-y^2$ orbital lies in the plane orthogonal to the O-Cu-N bond and provides
ferromagnetic coupling at the high temperatures. With decrease of temperature
the distortions of the half of the CuO$_5$N octahedra are changed, so that
two lobes of the $x^2-y^2$ orbital on the Cu1 sites turn out to be directed
along  the O-Cu-N bond. This leads to a strong antiferromagnetic coupling
and stabilization of the spin singlet (S=0) state in the half of the
exchange clusters at lower temperatures. As a result only a half of
local spins of the compound turns out to be unpaired and the value of observed effective
magnetic moment drastically drops.~\cite{Ovcharenko-2004,Gritsan-2010}

\section{Acknowledgments}
The authors are thankful to Prof. V.I. Ovcharenko for his useful comments.
This work is supported by the Russian Foundation for Basic Research
via  MK-3443.2013.2, RFFI-10-02-96011, RFFI-13-02-00374, RFFI-10-03-00075 and RFFI-10-02-00140,
the Ministry of education and science of Russia  (grants 12.740.11.0026 and 8436),
by the Ural branch of Russian
Academy of Science through the young-scientist program.
shown in this document. All the calculation were performed on the ``Uran'' cluster
of the IMM UB RAS.

\bibliography{../library}

\end{document}